\documentclass[twocolumn, pra,showpacs]{revtex4}

	\usepackage{amsmath}
	\usepackage{subfigure}
	\usepackage{epsfig}
	\usepackage[colorlinks,hyperindex]{hyperref}
	\hypersetup
	{
		colorlinks,%
		citecolor=black,%
		linkcolor=black,%
		urlcolor=black,%
	}

\makeindex

\begin{document}

\title{ Continuous Variable Cluster State Generation over the Optical Spatial Mode Comb }

\author{Raphael Pooser$^{1}$}
\email{pooserrc@ornl.gov}
\author{Jietai Jing$^{2}$}

\affiliation{$^{1}$Quantum Information Science Group, Oak Ridge National Laboratory, Oak Ridge, TN 37831}
\affiliation{$^{2}$State Key Laboratory of Precision Spectroscopy, East China Normal University, Shanghai 200062, China}


\begin{abstract}
One way quantum computing uses single qubit projective measurements performed on a cluster state (a highly entangled state of multiple qubits) in order to enact quantum gates. The model is promising due to its potential scalability; the cluster state may be produced at the beginning of the computation and operated on over time. Continuous variables (CV) offer another potential benefit in the form of deterministic entanglement generation.  This determinism can lead to robust cluster states and scalable quantum computation. Recent demonstrations of CV cluster states have made great strides on the path to scalability utilizing either time or frequency multiplexing in optical parametric oscillators (OPO) both above and below threshold. The techniques relied on a combination of entangling operators and beam splitter transformations. Here we show that an analogous transformation exists for amplifiers with Gaussian inputs states operating on multiple spatial modes. By judicious selection of local oscillators (LOs), the spatial mode distribution is analogous to the optical frequency comb consisting of axial modes in an OPO cavity. We outline an experimental system that generates cluster states across the spatial frequency comb which can also scale the amount of quantum noise reduction to potentially larger than in other systems.
\end{abstract}



\maketitle

\section{Introduction}
Quantum computation (QC) promises to solve factoring \cite{Shor}, database searches \cite{Grover}, and myriad optimization problems with dramatically greater efficiency than is possible in the classical computation model.  The traditional approach to quantum computation is based on unitary quantum logic gates that control the interactions between well-defined quantum states (aka qubits).  This approach is inherently difficult to scale because of challenges controlling decoherence. One way quantum computing \cite{BR2000} with continuous variables (CV) is potentially scalable due to the advantages provided by deterministic entanglement generation \cite{Lloyd}. A fault tolerance threshold for CV one way QC has also recently been discovered \cite{Menicucci1}.

Recent experiments exhibited the simplicity of one-way QC in the discrete variable regime by demonstrating Grover's algorithm with a four-qubit cluster state \cite{zeilinger2005} and by providing the first implementation of topological error correction with an eight photon cluster state \cite{RaussendorfNature2012}. However, these discrete implementations are more difficult to scale due to the probabilistic generation of entanglement necessary to form the cluster state.  Recent demonstrations of CV cluster states \cite{Zhang2006} have shown promise as the most scalable implementation of quantum resources for one way QC yet realized, with deterministic generation of entanglement that far outstrips the levels achieved by other systems.

Experimental demonstrations of CV cluster states have included 10000 time-multiplexed modes sequentially entangled (though only a few of these modes existed at any given time) \cite{Furusawa} and an experimental implementation of a cluster state in a frequency comb with more than 60 modes entangled and available simultaneously \cite{Olivier1}.
A cluster state with 16 simultaneous bright modes was also generated in multi-mode OPOs above threshold using pulsed light sources and filtered local oscillators (LOs) to measure entanglement among different portions of signal and idler pulses \cite{Fabre_time}. The elegance of these schemes, involving only beam splitter interactions and simple bipartite entanglement measurements, will likely see broad application in quantum optics and quantum information systems in the very near future.

The first theoretical proposal for multi-party CV entanglement in a compact single-OPO form utilized nonlinear crystals with concurrent nonlinearities \cite{Olivier2}, followed by experimental evidence that concurrent nonlinearities that could support such multipartite entanglement can be engineered \cite{Pooser}, and finally, the use of concurrences to generate clusters \cite{Pysher}. However, concurrent interactions in frequency  and polarization are a means to an end, and several other methods have also been proposed, including nonlinear interactions that are concurrent in time or space \cite{Fabre_space}. The main requirement is that the Bloch-Messiah reduction be applicable to the system \cite{Braunstein}; that is the combined nonlinear optics and linear optics systems can be described by linear Bogoliubov tranforms. The reduction states that any combination of multimode nonlinear interactions and interferometric interactions can be decomposed into an arrangement of single mode nonlinearities and linear optical elements, and it has been previously shown that this transform applies to CV cluster states derived from 2nd order  nonlinearities \cite{Weedbrook, Menicucci2}. In this manuscript, we outline the generation of mutipartite entanglement and CV cluster states utilizing concurrent nonlinear interactions spread across the spatial domain. We detail an experimental scheme using concurrent phase insensitive amplifiers based on four-wave mixing (FWM) in alkali metal vapors in which this method can be applied. Each concurrent amplifier operates on independent spatial modes. By choosing the LOs to measure specific entangled spatial modes, a spatial frequency comb can be generated from the amplified spatial modes, which can be mixed via a linear  transformation to generate a cluster state.

A fault tolerance threshold for CV QC in cluster states has also recently been derived in terms of quantum correlations below the shot noise \cite{Menicucci1}.
The FWM geometry has been shown to reach 9 dB of quantum noise reduction, and a cluster state with this level of squeezing would represent a promising step on the path to fault tolerant CV QC. The absence of an optical cavity in the FWM process allows for a compact and stable experiment that requires no phase locks, cavity length locks, or interferometric control, thereby enabling a potentially practical approach to quantum computation over cluster state resources.

\section{Optical spatial mode comb}
The ``optical spatial mode comb'' is analagous to the optical frequency comb. The requirements for generation of a comb involve an amplifier operating on a large continuum of modes followed by application of a filter to discretize the continuum (roughly speaking, and keeping in mind that a single frequency with infinitesimal bandwidth corresponds to a discrete monochromatic mode of the electric field, a single ``discrete'' axial mode in an optical cavity may consist of multiple monochromatic modes). In the case of a multimode OPO (either pulsed or CW), the resonant modes of an optical cavity that overlap with the phasematching bandwidth of a nonlinear medium are enough to define a comb. In the more familar case involving lasers, the pulse bandwidth may overlap the gain bandwidth and the axial modes of the optical cavity simultaneously as one means of making a comb. Naturally, optical cavities are narrow band spatial filters, with often only a few modes overlapping with the gain bandwidth (e.g.~a TEM$_{0,0}$ mode resonant inside a laser cavity). However, it is possible to use optical cavities to discretize continuous spatial frequencies into up to three simultaneously concurrent nonlinear interactions over spatial modes \cite{Fabre_space}. Without the discretizing filter, many nonlinear media emit into multiple spatial modes simultaneously. Perhaps the most important aspect is that the LO used for detection must match the desired modes, whether they be in the form of a comb or not. Here, we discuss using the input of a nonlinear amplifier to shape the LOs in such a way that an analogy to a frequency comb can be detected.

\begin{figure}[htb]
\includegraphics[width=3in]{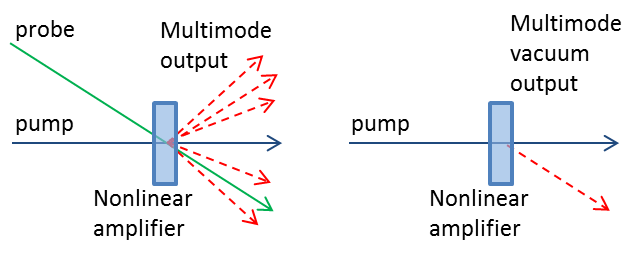}
\caption{A nonlinear amplifier with multimode vacuum output. Left: the amplifier generates a LO based on an amplified probe. Right: the amplifier mode corresponding the chosen LO. The LO will interfere with this vacuum mode during detection.}\label{LOgen}
\end{figure}

Consider a bare nonlinear medium, such as a BBO crystal, which emits photons into multiple spatial modes simultaneously. A biphoton spatial mode may be denoted by a pair of k-vectors, frequencies, polarizations, etc. In the limit of large gain, quantum correlations may be detected in any of the spatial modes as long as a LO matching them can be generated. The quickest route to generating the proper LO is to seed the same nonlinear process with a coherent field, treating it as an amplifier, to obtain a bright field whose phase front, frequency, polarization, etc., match the signal field \cite{BoyerScience} (see Fig.~\ref{LOgen}). Thus, the problem of measuring entanglement over the spatial mode comb is essentially a problem of producing the proper LOs. 
Here we assume that the amplifier is phase-insensitive; the requirement for generating LOs is that the pump and probe phases be set for amplification.

A large assumption implicit in Fig.~\ref{LOgen} is that an input field can be expressed in terms of an eigenfunction expansion of the amplifier modes. For instance, it would be difficult to input a probe field that has the same wavefront as an arbitrary Schmidt spatial mode \cite{schmidt}, but if the probe field can be expressed as an eignefunction expansion of the amplifier modes, then the generated LO will interfere with a summation of amplifier vacuum spatial modes. That is, if
$$\hat{a}_{LO}= \Sigma_{i=1}^n \alpha_i \hat{a}_i,$$
where $i$ corresponds to the $i^{th}$ spatial mode, then the overlap of the LO with some spatial distribution of output amplifier modes will be nonzero.
This can be empirically verified by ensuring a large nonlinear gain in the probe field, and is essentially determined by phase matching conditions. This summation of spatial modes may be considered orthogonal to a large number of other summations of spatial modes by showing that the separate sums do not overlap in their image planes.
One may produce multiple LOs as
\begin{equation}
\hat{a}_{LO_1} \hat{a}_{LO_2}= (\Sigma_{i=1}^n \alpha_i \hat{a}_{i})(\Sigma_{j= 1}^m \beta_j \hat{a}_{j}). \label{LOs}
\end{equation}
Assuming that each mode is independently amplified by the nonlinear medium, the interaction Hamiltonian for a single amplifier mode is (assuming $\chi^{(2)}$ media)
\begin{equation}
H = i\hbar\chi^{(2)} a_{s,k_s} a_{i,k_i} a^\dagger_{p,k_p} + H.C., \label{H1}
\end{equation}
where each $k$ corresponds to the k-vector for a given optical frequency (signal, s, or idler, i). We may consider that a single amplifier mode consists of a set of three k-vectors that satisfies the phase matching condition $k_s+k_i-k_p=0$ (in the limit of exact phase-matching). Many amplifier modes may be concurrently phase matched, such that one has a set of independent interaction Hamiltonians (assuming equal gain for each mode):
\begin{equation}
H = i\hbar\chi^{(2)} \sum a_{i,k_i} a_{j,k_j} a^\dagger_{p,k_p} + H.C. \label{H2}
\end{equation}
Solving the equations of motion yields a set of biparite entangled modes (for $i\neq j$) whose entanglement witnesses are the bipartite EPR operators \cite{ReidEPR}:
\begin{eqnarray}
\hat{X}_i(t)-\hat{X}_j(t) = (\hat{X}_i(0)-\hat{X}_j(0))e^{-\kappa t} \label{EPRops1} \\ \label{EPRops2}
\hat{P}_i(t)+\hat{P}_j(t) = (\hat{P}_i(0)+\hat{P}_j(0))e^{-\kappa t}
\end{eqnarray}
where the pump field operater $a_p$ has been approximated as a classical number and subsumed in $\kappa$.
We note that the form of Eqs.~\ref{H2} - \ref{EPRops2} implies that the amplifier modes denoted by $k_i$, $k_j$ are coincident with the squeezed eigenmodes of the system \cite{Bennink}. That is, the interaction Hamiltonian has been written with a  squeezing parameter matrix, $H=i\hbar \kappa~\boldsymbol{C} \cdot \boldsymbol{A} + H.C.$, where $\boldsymbol{A}$ is a field mode operator vector and $\boldsymbol{C}$ is an interaction matrix which defines a graph of mode pairs (for the case of four modes for brevity):
\begin{equation}
\boldsymbol{C}=i\hbar \kappa
\left( \begin{array}{cccc}
0 & 1 & 0 & 0 \\
1 & 0 & 0 & 0 \\
0 & 0 & 0 & 1 \\
0 & 0 & 1 & 0 \end{array} \right)
\end{equation} \label{opmatrix}

\begin{figure}[!h]
\includegraphics{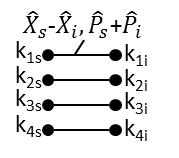}
\caption{Bipartite graphs for eight entangled modes. Each node represents a k-vector for either the signal or idler, while each edge represents an entangling interaction. One of the entanglement witnesses is noted.}\label{bigraph}
\end{figure}

The corresponding graph states for eight modes are given in Fig.~\ref{bigraph}.
Note that while each k-vector pair represents an individual mode of the amplifier, the concept of the mode is immaterial until measured with a LO. One may produce two LOs, $k_+=k_{1s}+k_{2s}+k_{3s}+k_{4s}$, $k_-=k_{1i}+k_{2i}+k_{3i}+k_{4i}$ per Eq.~\ref{LOs} to detect the modes in Fig.~\ref{bigraph} with two homodyne detectors (one corresponding to each LO), which effectively transforms the graphs to a single bipartite graph. The converse is also true: one may produce arbitrary spatial mode combs by selecting appropriate LOs within the amplifier's phase matching bandwidth in order to detect individual k-vectors or combinations of k-vectors analogous to those in Fig.~\ref{bigraph}.

\section{Dual rail cluster states with spatial nodes}
\begin{figure}[!h]
\includegraphics[width=2.5in]{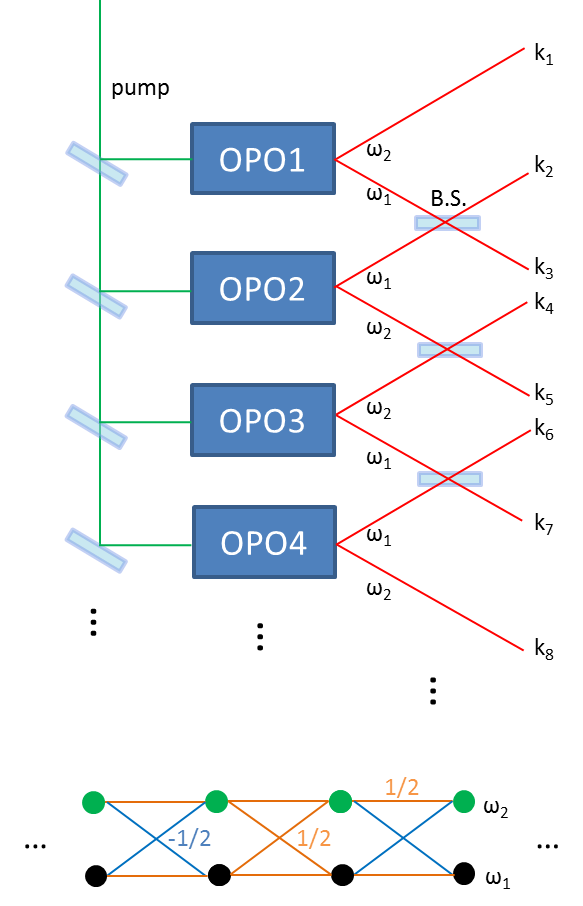}
\caption{An unfolded OPO with multiple concurrent nonlinearities can be treated as a system of cascaded OPOs with single nonlinearites (top). In this picture, each OPO emits into distinct spatial modes. The interference of each half an EPR state on balanced beam splitters yields the dual rail cluster state (bottom), with weighted edges $-\frac{1}{2}$ (blue lines) and $\frac{1}{2}$ (orange lines).} \label{concOPO}
\end{figure}

It was previously shown that EPR states (eigenstates of the operators in \ref{EPRops1} and \ref{EPRops2}), which are cluster states of the type in Fig.~\ref{bigraph}, can be concatenated into a dual rail cluster state, or quantum wire \cite{Menicucci2, Furusawa, Olivier1}. While the proposal and implementations have been in compact, single OPOs, it is illuminating to draw a more explicit equivalent picture via Bloch-Messiah reduction. Unfolding the dual-pumped, single OPO cavity in \cite{Olivier1} into a series of OPOs, one can show that \textit{identical} two mode squeezers interfered on a train of beam splitters leads to a dual rail cluster state (after applying a $-\pi/2$ phase shift to every odd mode, or a redefinition of the quadrature operators), as shown in Fig.~\ref{concOPO}.

This picture is analogous to that drawn over the frequency comb in optical cavities. The differences are in the degree of freedom used to represent the nodes on each graph. In the axial mode case, each node is an axial mode of an optical cavity, separated by a free spectral range, where two or more pump fields serve to overlap optical frequencies with an additional degree of freedom (such as polarization) in order to allow for interference of distinct modes later (the axial mode pairs that comprise the EPR states would otherwise not interfere with one another due to their frequency separation). In the spatial mode case, the k-vector serves as a stand in for the cavity axial modes. Since the superpositions of spatial modes are separable after diffraction limited propagation, an additional degree of freedom is not needed in order to interfere distinct modes at beam splitters and subsequently homodyne detectors. The entanglement witnesses that need to be measured to verify the bipartite correlations, between the first two modes for instance, necessary to form a cluster state are \cite{Furusawa}:
\begin{eqnarray}
\left[(X^{(2)}_1+X^{(3)}_1) - (X^{(4)}_2-X^{(5)}_2)\right]e^{-\kappa t}\label{wit1} \\ 
\left[(P^{(2)}_1+P^{(3)}_1)-(P^{(4)}_2-P^{(5)}_2)\right]e^{-\kappa t} \label{wit2}
\end{eqnarray}
The subscripts in Eqs.~\ref{wit1}-\ref{wit2} denote the frequency in Fig.~\ref{concOPO} while the superscripts denote the k-vector.

\section{Experimental implementation}
\begin{figure}[!h]
\includegraphics{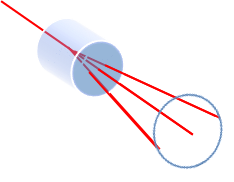}
\caption{A cylindrical alkali vapor cell with pump beam propagating through the center and bipartite emission into two fields symmetric about the pump field. The circle denotes a line on which all signal k-vectors experience the same gain if the pump field creates a cylindrically symmetric gain region. A Gaussian shaped pump would result in such a gain region.}\label{cyl}
\end{figure}

Here we outline a multispatial mode amplifier configuration that yields a dual rail cluster state over the optical spatial mode comb. The scheme uses a well known nonlinear amplifier: FWM in alkali vapor based on a double $\Lambda$ system near the D1 \cite{McCormick, cesium} or D2 \cite{Pooser2} transition. The amplifier is based on a third order nonlinearity which is isotropic in homogeneous vapor. A finite interaction length quasiphasematches a set of k-vectors that fall within an angular acceptance bandwidth \cite{BoyerScience}. The inverse transverse gain region sets the size of a ``spatial mode'' in the far field \cite{Brambilla}, otherwise known as a coherence area \cite{brambilla_lugiato_coh_area}. These coherence areas can be considered independent spatial modes in the sense that they do not interfere in their image planes and can be detected with separate homodyne detectors as discussed in section II. The amplifier can be made analogous to our hypothetical amplifier with equal gain for all modes by considering one set of modes along a circle within the angular acceptance bandwidth (the gain is cylindrically symmetric about the gain region, given the gain region's own cylindrical symmetry; see Fig.~\ref{cyl}).

\begin{figure}[htb]
\includegraphics[width=3in]{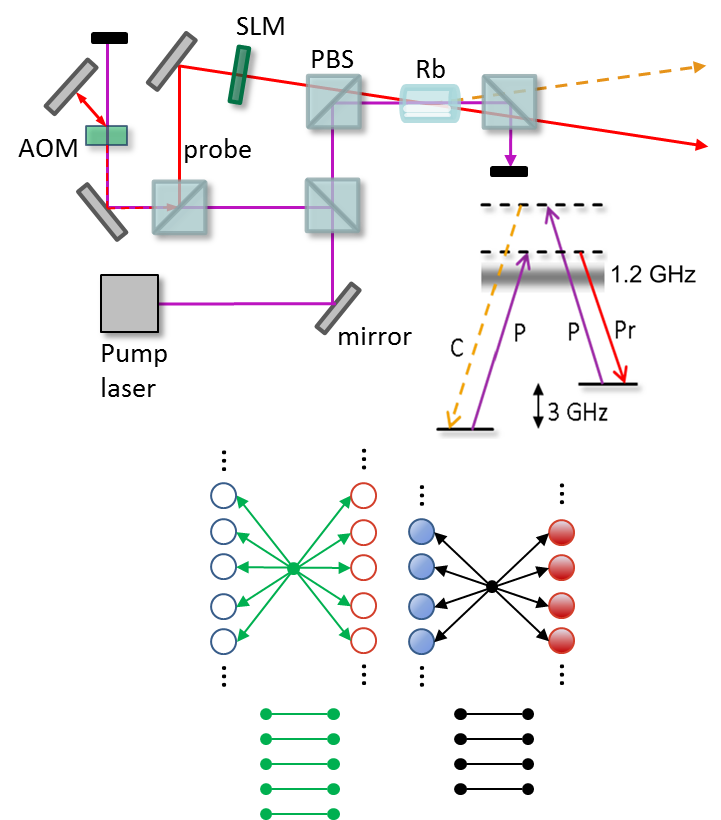}
\caption{Experimental setup for a multimode amplifier based on FWM. The probe field is derived from the pump field at an acousto-optic modulator (AOM) before being shaped by the spatial light modulator (SLM) into a series of ``dots''. The energy level diagram shows a double $\Lambda$ system in Rb vapor at the D1 line (795nm). Bottom: output modes for the probe (red) and conjugate (blue) fields for two pump positions within the vapor cell for a given input probe image. The black and green arrows connecting closed and open circles respectively denote nonlinear interactions between mode pairs that are correlated in an image reflected symmetrically about the pump axis. Each line carries equal weight if the input probe image is coincident with a semicircular gain region symmetric about the pump axis in the far field. In that case, the graph states shown in the lower portion correspond to the output EPR states for each mode pair.}\label{setup}
\end{figure}

The Hamiltonian for the fields along the constant gain circle is given by
\begin{equation}
H = i\hbar\chi^{(3)} \Sigma a_{i,k_i} a_{j,k_j} a^\dagger_{p_1,k_{p_1}}a^\dagger_{p_2,k_{p_2}} + H.C.
\end{equation}
The process is a third order nonlinearity supported by the double $\Lambda$ system shown in Fig.\ref{setup}. The two pump fields can be taken as classical numbers, which reduces the system to an effective second order interaction with EPR eigenstates as in Eqs.~\ref{EPRops1}, \ref{EPRops2}.

Using a spatial light modulator (see Fig.~\ref{setup}), multiple probe fields can be shaped from a single beam in the form of an image input to the amplifier. Probe fields on one half of the output circle produce conjugate fields on the opposite half of the circle, as shown in the bottom half of Fig.~\ref{setup}. If the squeezing parameters are equal between each mode pair, and if each ``dot'' in Fig~\ref{setup} does not interfere with any other ``dot'' in its image plane, then the entanglement witnesses \ref{wit1}-\ref{wit2} apply directly, where each ``dot'' is a single mode at one of two frequencies.

The caveat in Fig.~\ref{setup} is that certain practical parameters of the experiment must be taken into account. Only a finite number of independent modes will fit within the constant gain region (primarily dictated by the pump focus). For instance, it has been shown that about 200 modes can fit within the gain bandwidth for typical pump/probe beam sizes of 1000~$\mu$m / 500~$\mu$m respectively, with detunings of 1~GHz / 3036~MHz, and a relative incidence angle of 3~mrad~\cite{BoyerScience}. The modes were counted by observing the transverse conjugate beam size within the far field as a function of probe focus for a constant pump focus. A separate experiment measured the approximate size and physical arrangement of coherence areas within an image produced by the FWM medium as a function of pump/probe overlap~\cite{Lawrie_compOE}. These empirical observations show that it is possible to amplify and separate a large number of coherence areas within a single FWM process.

\begin{figure}[htb]
\includegraphics[width=3in]{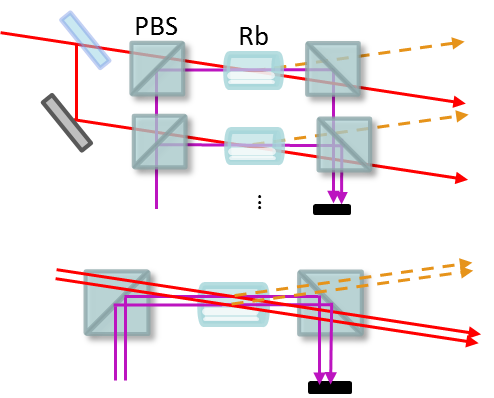}
\caption{Top: cascaded 4WM setup. Each vapor cell uses an identical pump and probe to amplify a set of LOs. Phase control must be maintained among every beam, and the vapor cells must have identical gain profiles. Bottom: a single vapor cell is spatially multiplexed. Each spatial region uses an identical pump and probe. }\label{cascades}
\end{figure}

\begin{figure}
\vspace{0.1in}
\includegraphics[width=2in]{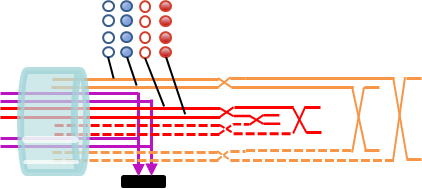}
\caption{Schematic interference of four LO combs with vacuum modes. Each LO pair, denoted by the orange and red lines, interferes with its corresponding frequency from the other pump. The vacuum modes (dotted lines) from two separate pumps interfere with each other in the same manner on the same beam splitters, spatially separated from the LOs so as not to interfere with them. Thus, the first set of four beam splitters may be condensed to two in the experimental setup. In the second set of beam splitters the LOs interfere with the corresponding vacuum modes. Each line terminates at a set of balanced detectors. The same beam splitter can again be used for an entire spatial comb, provided that each coherence area can be sent to separate photodiodes after (16 photodiodes would be needed in this example).}\label{combs}
\end{figure}

In order to scale the number of modes indefinitely, multiple pumps and multiple probes can be used, as shown in Fig.~\ref{cascades}. This can be accomplished by either splitting pump and probe fields and using multiple spaces in the same vapor cell, or by cascading multiple vapor cells. The latter option has an added advantage in that it will allow for cascaded gain regions which can lead to increased squeezing \cite{jietai}. Its similarity with Fig.~\ref{concOPO} is also apparent. The former has the advantage that phase control is much simpler, as the pump and probe fields may be split as close to the cell as possible, ensuring that the optical paths are as close to one another as possible.  Each beam in Fig.~\ref{cascades} may be taken to represent LOs for either a single or multiple k-vectors (such as those produced with images as shown in Fig.~\ref{setup}). The fields can be interfered with one another if the gain regions are identical. It has been shown empirically that the interference of whole images with their local oscillators yields high visibility even for complicated arrangements in which the LOs are amplified and attenuated in multiple optical paths~\cite{Alberto_nature, pooser3}, rather similar to the arrangement proposed in Fig.~\ref{combs}

Finally, modes with like frequency must be interfered on beam splitters in order to concatenate their graphs. The LOs must undergo the same process in order to ensure good mode matching to the vacuum fields during the detection process, thus, both the vacuum and the LOs must be interfered on beam splitters amongst themselves. Then, the LOs and vacuums are finally mixed during the detection stage.

A practical question arises from these requirments. It is natural to consider the question of how many beam splitters would be required to interfere all coherence areas with each other. If the number of beam splitters is approximately equivalent to the number of modes, then the scheme is perhaps less practical than other schemes which may use a single beam splitter multiplexed in time or optical frequency. Fortunately, we may likewise multiplex beamsplitters spatially by treating each spatial mode comb as an image. Entire images, which contain multiple coherence areas, can then be interefered with one another. This concept was verified empirically in ~\cite{Alberto_nature, pooser3} while the frequency-independent separability of coherence areas was empirically studied in~\cite{Lawrie_compOE}. Thus, a single beamsplitter is needed in order to concatenate two spatial mode combs and a second beam splitter is needed for the homodyne detection step. A single beam splitter can be used to concatenate multiple spatial mode combs by further multiplexing a single beam splitter spatially. The limit to the number of modes that can be interfered on a single beam splitter is essentially dictated by the number of spatial modes that can be imaged independently in each port.

Fig.~\ref{combs} shows a schematic setup of the complete system. Two pump fields generate four LO combs using an input image on two probe fields, where both input sets are derived from the same initial field (generally, N pump fields with N probes may produce 2N LOs with a number of coherence areas, where the N input sets are all derived from the same initial field).
Because the frequency combs can be treated as individual images, they can interfere with one another on the same beam splitter with high visibility. That is, the ``dots'' in Fig.~\ref{setup} can be treated as a single image and interfere with one another both during the concatenation step and at the homodyne detector where the LOs and vacuum fields interfere. Afterwards, the coherence areas are finally separated and sent to individual photodiodes for balanced detection. In the limit of perfect LO-signal overlap, the entanglement witnesses can be used to verify the final state as the dual rail cluster state shown in Fig.~\ref{concOPO} \cite{ent_witness}.

\subsection{Multimode homodyne detection}
The LOs undergo the same interferences and traverse the same optical paths as the two mode squeezed vacuum signals, meaning their wavefronts will interfere well with the signals. However, each independent vacuum mode is actually an expansion of amplifier modes. If all modes can be detected, then measuring the bipartite EPR operators should show squeezing between mode pairs. In the limit that the LOs are perfectly matched to the signals, the noise on the position difference operator (chosen for brevity, but other entanglement witnesses follow similarly) approaches that of two individual entangled modes (normalized to the shot noise):
\begin{equation}
\Delta x_-^2 =
 1 + 2 \eta (G-1 - \sqrt{G(G-1)}),
\end{equation}
where $\eta$ is the detector efficiency and $G$ is the nonlinear gain.

However, if the LOs are misaligned from the correct vacuum modes, the detected correlations are reduced:
\begin{equation}
\begin{split}
\Delta x_-^2 =\frac{P_0}{P_{\text{tot}}} \left(1+2 \eta _d( G-1- \sqrt{G(G-1)}\right) + \\
\sum _{i=1}^N \frac{P_i}{P_{\text{tot}}}\left(1 + 2 \eta_i (G-1 - \sqrt{G(G-1)} \right)+ \\
\sum _{i=1}^N \frac{\left(P_{\text{tot}}-P_0-P_i\right)}{P_{\text{tot}}}\left(1+2 \eta _i(G-1) \right),
\end{split} \label{LO_overlap}
\end{equation}
where $P_{tot}$ is the total power contained in the LO, $P_0$ is the portion of the LO that wholly overlaps with the corresponding spatial modes in the vacuum field, $\eta_d$ is the detector efficiency, and $P_i$ is the portion of the LO power that overlaps partially with the $i^{th}$ vacuum spatial mode, where the overlap is determined by an effective reduced detector efficiency $\eta_i < \eta_d$. A typical number for $\eta_d$ is 95\%-96\% in off the shelf components (for which the FWM system based on Rb is capable of 9dB of quantum noise reduction), while custom photodiode coatings can achieve efficiencies of greater than 99\%. Also note the constraint on the total power:
\begin{equation}
\sum _{i=1}^N P_i=P_{\text{tot}}-P_0,
\end{equation}
which implies that $\sum _{i=1}^N P_i \rightarrow 0$ as $P_0 \rightarrow P_{tot}$.
The third term in Eq.~\ref{LO_overlap} is due to the excess noise contained within uncorrelated vacuum spatial modes that are accidentally detected in each LO. Note that the nonlinear interaction spontaneously amplifies all vacuum amplifier modes, and the LOs are used to pick out each mode in a comb. If the LOs pick out neighboring modes, they will measure anticorrelations. This is a disadvantage relative to an OPO, whose misaligned LO measurement would yeild only the first two terms in Eq.~\ref{LO_overlap} because the optical cavity effectively filters out all of the vacuum modes that do not overlap with the LO. All of the entanglement witnesses required to measure the cluster state will suffer from this excess noise, meaning that the purity of the cluster state will be degraded by LO misalignment. Very good alignment of the LOs with their signal modes will minimize this effect, as it requires $P_0=P_{\text{tot}}$.

\subsection{Practical considerations}
Another consideration is the initial state purity before homodyne detection. As cluster states are resources for one-way quantum computers, which run quantum algorithms that formally start from initialized pure states, quantum computing with statistical mixtures is not necessarily defined. Therefore it is important to consider whether a system that produces entanglement resources is also capable of producing pure states. We note that the FWM process is under certain conditions a quantum-noise-limited amplifier~\cite{pooser3}. This means that the noise added to a pure input state is the minimum amount required by quantum mechanics under these conditons, and the two-mode output is in a pure two-mode squeezed state. This is true if all other classical noise sources on the input can be minimized, meaning the probe is in a coherent state. This can be achieved by minimizing classical laser noise at the working analyzer frequency.

Finally, a physical system that can produce cluster states would not be useful for universal quantum computing without supporting a non Gaussian operation~\cite{Lloyd}. A cubic phase gate is one operation that would fulfill this requirement~\cite{preskill2001}. We note that a large amount of quantum noise reduction is needed to achieve high cubic phase gate fidelity. The FWM system is potentially capable of the levels of squeezing needed for a successful application, but we relegate further discussion of non Gaussian operations to a future study.

\section{conclusion}
In this manuscript we have drawn an analogy between the optical spatial mode comb and the optical frequency comb over which dual rail cluster states can be generated. We presented an example of an experimental system in which these cluster states can be generated and detected using images to synthesize appropriate LOs. The experimental system considered suffers a potential disadvantage with respect to the single OPO implementations, in that excess noise is introduced for any LO misalignment. However, the system offers the potential advantages of simple phase control, ease of alignment, and scalability via the use of multiple gain regions.

\begin{acknowledgements}
We would like to thank Olivier Pfister and Pavel Lougovski for useful discussions.
This work was performed at Oak Ridge National Laboratory, operated by UT-Battelle for the U.S. Department of energy under contract no.~DE-AC05-00OR22725. J. J acknowledge the support from the NSFC under Grants No. 11374104 and No. 10974057, the SRFDP (20130076110011), the Shu Guang project(11SG26), the Program for Eastern Scholar, and the NCET Program(NCET-10-0383).
\end{acknowledgements}


\begin{thebibliography}{}
\bibitem{Shor}
P.~W.~Shor,
Proc. 35nd Annual Symposium on Foundations of Computer Science, IEEE Computer Society Press, Washington, DC, 124-134 (1994).
\bibitem{Grover}
L.~K.~Grover,
Phys. Rev. Lett. \textbf{79}, 325 (1997).
\bibitem{BR2000}
R.~Raussendorf and H.~J.~Briegel, Phys. Rev. Lett. \textbf{86}, 5188 (2001).
\bibitem{Lloyd}
S.~Lloyd and S.~L.~Braunstein,
Phys. Rev. Lett. \textbf{82}, 1784 (1999).
\bibitem{Menicucci1}
N.~C.~Menicucci,
Phys. Rev. Lett. \textbf{112}, 120504 (2014).
\bibitem{zeilinger2005}
P.~Walther \emph{et al.},
Nature \textbf{434}, 169-176 (2005).
\bibitem{RaussendorfNature2012}
X-C.~Yao \emph{et al.},
Nature \textbf{482}, 489-494 (2012).
\bibitem{Zhang2006}
J.~Zhang and S.~L.Braunstein, Phys. Rev. A \textbf{73}, 032318 (2006).
\bibitem{Furusawa}
S.~Yokoyama \emph{et al.},
Nature Photonics \textbf{7}, 982-986 (2013).
\bibitem{Olivier1}
M.~Chen,  N.~C.~Menicucci, and O.~Pfister,
Phys. Rev. Lett. \textbf{112}, 120505 (2014).
\bibitem{Fabre_time}
J.~Roslund,	 R.~M. de Ara��jo,	 S.~Jiang,	 C.~Fabre and N.~Treps,
Nature Photonics, \textbf{8}, 109 (2013).
\bibitem{Olivier2}
O.~Pfister, S.~Feng, G.~Jennings, R.~Pooser, and D.~Xie, Phys. Rev. A \textbf{70}, 020302 (2004).
\bibitem{Pooser}
R.~C.~Pooser and O.~Pfister,
Opt. Lett. \textbf{30}, 2635 (2005).
\bibitem{Pysher}
M.~Pysher, Y.~Miwa, R.~Shahrokhshahi, R.~Bloomer, and O.~Pfister,
Phys. Rev. Lett. \textbf{107}, 030505 (2011).
\bibitem{Fabre_space}
B.~Chalopin, F.~Scazza, C.~Fabre, and N.~Treps, Phys. Rev. A \textbf{81}, 061804(R) (2010).
\bibitem{Braunstein}
S.~L.~Braunstein, Phys. Rev. A \textbf{71}, 055801 (2005).
\bibitem{Weedbrook}
P.~van Loock, C.~Weedbrook, and M.~Gu,
Phys. Rev. A \textbf{76}, 032321 (2007);

N.~C.~Menicucci, S.~T.~Flammia, H.~Zaidi, and
O.~Pfister,
Phys. Rev. A \textbf{76}, 010302
(2007);

N.~C.~Menicucci, X.~A.~Ma, and T.~C.~Ralph,
Phys. Rev. Lett. \textbf{104}, 250503 (2010);
\bibitem{Menicucci2}
N.~C.~Menicucci,
Phys. Rev. A \textbf{83}, 062314 (2011).
\bibitem{BoyerScience}
V.~Boyer, A.~M.~Marino, R.~C.~Pooser, and P.~D.~Lett,
Science \textbf{321}, 544 (2008).
\bibitem{schmidt}
C.~K.~Law and J.~H.~Eberly,
Phys. Rev. Lett. \textbf{92}, 127903 (2004).
\bibitem{ReidEPR}
M.~D.~Reid,
Phys. Rev. A \textbf{40}, 913 (1989).
\bibitem{Bennink}
R.~S.~Bennink and R.~W.~Boyd, Phys. Rev. A \textbf{66}, 053815 (2002).
\bibitem{McCormick}
C.~F.~McCormick, V.~Boyer, E.~Arimondo, and P.~D.~Lett,
Opt. Lett. \textbf{32}, 178 (2007);

C. Liu, J. Jing, Z. Zhou, R.~C.~Pooser, F. Hudelist, L. Zhou, and W. Zhang,
Opt. Lett. \textbf{36}, 2979 (2011);

\bibitem{cesium}
M.~Guo \emph{et al.}
Phys. Rev. A \textbf{89}, 033813 (2014).
\bibitem{Pooser2}
R.~C.~Pooser, A.~M.~Marino,~V. Boyer, K.~M.~Jones, and P.~D.~Lett,
Opt. Exp. \textbf{17}, 16723 (2009).
\bibitem{Brambilla}
E. Brambilla, A. Gatti, M. Bache, and L. A. Lugiato, Phys. Rev. A \textbf{69}, 023802 (2004).
\bibitem{brambilla_lugiato_coh_area}
O.~Jedrkiewicz \emph{et al.},
Phys. Rev. Lett. \textbf{93}, 243601 (2004).
\bibitem{Lawrie_compOE}
B.~J.~Lawrie and R.~C.~Pooser, Optics Express, \textbf{21} 7549 (2013). 
\bibitem{jietai}
Z. Qin, L. Cao, H. Wang, A.~M.~Marino, W. Zhang, and J. Jing,
Phys. Rev. Lett. \textbf{113}, 023602 (2014).

\bibitem{Alberto_nature}
A.~M.~Marino, R.~C.~Pooser, V.~Boyer, and P.~D.~Lett, Nature \textbf{457} 859 (2009).

\bibitem{pooser3}
R.C. Pooser, A.M. Marino, V. Boyer, K.M. Jones, and P.D. Lett, Phys. Rev. Lett. \textbf{103}, 010501 (2009)

\bibitem{ent_witness}
P.~van Loock and A.~Furusawa, Phys. Rev. A \textbf{67}, 052315 (2003);

N. C. Menicucci, S. T. Flammia, and P. van Loock, Phys. Rev. A \textbf{83}, 042335 (2011);

M. Gu, C. Weedbrook, N. C. Menicucci, T. C. Ralph,
and P. van Loock, Phys. Rev. A \textbf{79}, 062318 (2009).


\bibitem{preskill2001}
Daniel Gottesman, Alexei Kitaev, and John Preskill, Phys Rev A, \textbf{ 64}, 012310 (2001)



\end{thebibliography}
\end{document}